# Phase separation of saturated micellar network and its potential applications for nanoemulsification

Mihail T. Georgiev, Lyuba A. Aleksova, Peter A. Kralchevsky*, Krassimir D. Danov

*Department of Chemical and Pharmaceutical Engineering, Faculty of Chemistry and Pharmacy, Sofia University, Sofia 1164, Bulgaria*

ORCID Identifiers: Peter A. Kralchevsky: 0000-0003-3942-1411 ; Krassimir D. Danov: 0000-0002-9563-0974

**Abstract**

Phase separation of saturated micellar network, as a result of cross-linking of branched micelles is established in mixed solutions of the anionic surfactant sodium laurylethersulfate (SLES) and the zwitterionic cocamidopropyl betaine (CAPB) in the presence of divalent counterions: $Ca^{2+}$, $Zn^{2+}$ and $Mg^{2+}$. The saturated network appears in the form of droplets, which are heavier than water and sediment at the bottom of the vessel. In the case of $Mg^{2+}$, the sedimented drops coalesce and form a separate multiconnected micellar phase – a supergiant surfactant micelle. For this phase, the rheological flow curves show Newtonian and shear-thinning regions. The appearance/disappearance of the Newtonian region marks the onset of formation of saturated network. The addition of small organic molecules (fragrances) to the multiconnected micellar phase leads to an almost spontaneous formation of oil-in-water nanoemulsion. The nanoemulsification capacity of the multiconnected micellar phase decreases with the rise of the volume of the oil molecule. A possible role of the network junctions in the nanoemulsification process can be anticipated. The properties of the multiconnected micellar phases could find applications in extraction and separation processes, in drug/active delivery, and for nanoemulsification at minimal energy input.

___________________________________________________________________________________

* Corresponding author. Fax: +359 2 9625643.
  Email address: pk@lcpe.uni-sofia.bg (P.A. Kralchevsky)

# 1. Introduction

The plot of viscosity of a micellar solution with ionic surfactant vs. the concentration of added salt usually has a typical shape with high maximum, known as "salt curve" [1-5]. This behavior is explained with the initial growth and entanglement of wormlike micelles, followed by a transition to branched micelles. In other words, the entangled network to the left of the maximum is replaced with an unsaturated branched network to the right of the maximum. With the rise of salt concentration, the degree of crosslinking in the branched network increases, until eventually a saturated network of multiconnected wormlike micelles is formed (Fig. 1); see Refs. [6-13]. To the right of the peak, the decrease of viscosity with the increase of salt concentration, which has been observed in many experimental studies [1-5,14], can be explained with the formation of mobile junctions that can slide along the linear micelles [8,12,15,16]. The formation of branched micelles and multiconnected network (as a limiting case) has been used as explanation of the rheological and phase behavior of numerous experimental systems [9-11,17-23]. The viscosity curve can be essentially affected by the addition of fatty acids, alcohols, and fragrances [5,24-29].

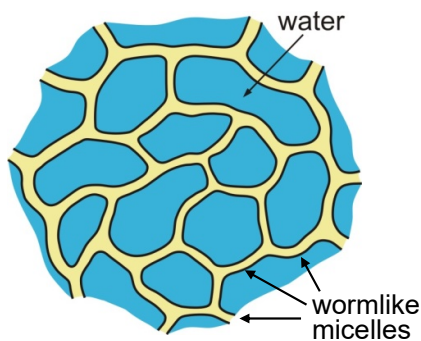

**Fig. 1.** Part of a saturated network of multiconnected wormlike surfactant micelles; after Ref. [9]. This is a simplified two-dimensional sketch – the real network is three-dimensional.

Phase separation of isotropic micellar phase was first observed by Appel and Porte [6] in solutions of cationic surfactants, cetylpyridinium bromide and cethyltrimethylammonium bromide, with added electrolytes, $NaClO_3$ and $NaNO_3$. In a subsequent study [7], it was suggested that a possible explanation of the observed "L1/L1 phase separation" can be the formation of an infinite cluster of branched cylinders. (L1 denotes isotropic micellar phase.) By theoretical analysis of the occurrence of crosslinks in semi-dilute solutions of giant wormlike micelles, Drye and Cates [8] arrived at the conclusion that interconnection of all wormlike micelles into a large saturated network is possible and this would lead to phase separation of the micellar network. The formation of such network is related to the disappearance of the endcaps of the wormlike micelles, which become energetically unfavorable under certain conditions (e.g. high salt concentrations for ionic micelles) and their transformation into junctions in a branched structure. (In this respect, Fig. 1 visualizes the topology of the micellar network, rather than its stereometry.) The formation of multiconnected aggregate is the final state in the structural transformation of the linear ionic surfactant aggregates upon the rise of salt concentration: wormlike micelles → branched micelles → saturated network [8-12,30]. With the development of the cryogenic transmission electron microscopy (cryo-TEM), observations of branched micelles and large multiconnected domains have been reported for various systems [31-40].

It should be noted that formation of branched micellar aggregates and saturated networks has been observed not only with ionic surfactants at high salt concentrations, but also with nonionic surfactants, such as polyoxyethylene alkyl ethers, upon the rise of temperature (close to the cloud point) [41-43], as well as with alkyl glucosides with the rise of surfactant concentration [44].

As already mentioned, the interconnection of wormlike micelles into a large saturated network should lead to its phase separation [6-8,13,45-47]. Such phase separation has been reported only in few experimental studies as an element of experimental phase diagrams and/or as multiconnected microdomains observed by cryo-TEM [6,15,40,48,49]. Systematic experimental studies on the rheological, electrophoretic, solubilization and other properties of the phase-separated saturated network, as well as on its potential applications are needed.

Here, we report that phase separation of saturated network as a multiconnected micellar phase occurs in mixed solutions of anionic and zwitterionic surfactants in the presence of divalent counterions. The surfactants are sodium laurylethersulfate with one ethylene oxide group (SLES) and cocamidopropyl betaine (CAPB), which find broad applications in shampoos and other personal care products [3]. The counterions are $Ca^{2+}$, $Zn^{2+}$ and $Mg^{2+}$. The yield of separated multiconnected micellar phase is the greatest in the case of $Mg^{2+}$ – it is possible to produce macroscopic amounts of this phase and to investigate its properties in various experiments. This phase is transparent, of viscosity 300–600 mPa·s. It is heavier than water, so that it is separated as sediment at the bottom of the vessel. When poured from one to another vessel, this phase is prone to produce long and relatively stable liquid fibers, i.e. it possesses elongational elasticity. An interesting property of the saturated micellar network is that in contact with small organic molecules (fragrances) it is transformed into oil-in-water nanoemulsion of drop diameter 100 – 200 nm. The latter finding implies that the multiconnected micellar phase could find applications for the production of nanoemulsions with their broad spectrum of pharmaceutical and biomedical applications for drug delivery [50-56]. Self-assembled structures, like the cubosomes and hexosomes, have been reported to have analogous potential applications [57-59].

The present article is our first report on the obtaining and properties of the multiconnected micellar phase from CAPB + SLES in the presence of divalent counterions. In Section 2, the used materials and methods are described. In Section 3, we report the similarities and dissimilarities between the systems with $Ca^{2+}$, $Zn^{2+}$ and $Mg^{2+}$ counterions. In Section 4, the bicontinuous structure of the separated phase is confirmed by using lipophilic and hydrophilic dyes. The effects of CAPB/SLES ratio and $MgCl_2$ concentration are investigated. The zeta-potential of droplets from the saturated network is determined. In Section 5, the effect of $MgCl_2$ concentration on the rheology of the multiconnected micellar phase is investigated and the ability of this phase to rebuild its structure after intensive shearing is examined. Finally, Section 6 is dedicated to the nanoemulsification of oils (fragrances) in contact with the saturated network. It should be noted that a complete systematic investigation on the properties of this interesting system goes beyond the volume of a standard paper and should be continued in subsequent studies.

Concerning the terminology, the phase separation of saturated network was initially called "L1/L1 phase separation" [7]. In the next studies, the terms "saturated network" [8,9] and "multiconnected network of wormlike micelles" [10,15] were introduced and became generally accepted. In this study, the terms "saturated network" and "multiconnected micellar phase" are used as synonyms.

**2. Materials and methods**

*2.1. Materials*

The used anionic surfactant has been sodium laurylethersulfate with one ethylene oxide group (SLES-1EO or more briefly SLES), a product of Stepan Co; trade name STEOL CS-170; molecular mass 332.4 g/mol. Its critical micellization concentration (CMC) in pure water is $7\times10^{-4}$ M. The used zwitterionic surfactant has been cocamidopropyl betaine (CAPB) with a quaternary nitrogen atom and a carboxylic group (Fig. 2), product of Evonik Industries, under the trade name Tego®Betain F50; molecular weight 356 g/mol. The CMC of CAPB in pure water is $9 \times 10^{-5}$ M. Because the molecular weights of the two surfactants are not so different, the mol/mol composition of their mixed solutions is close to the weight/weight composition. In our experiments, the pH of the solutions was in the range 5 – 6, so that CAPB is in its zwitterionic form.

The used salts, $CaCl_2$, $ZnCl_2$ and $MgCl_2$ were products of Merck (Darmstadt, Germany). The structural formulas of the two surfactants and of the four fragrances, viz. linalool, citronellol, limonene, and benzyl salicylate, used in the solubilization/ nanoemulsification experiments are shown in Fig. 2.

Table 1 shows data for some properties of the above fragrances. Their lipophilicity is characterized by log$P$, where $P$ is the partition coefficient of the fragrance between octanol and water. The van der Waals volume of a molecule is defined as the space occupied by the molecule, which is impenetrable to other molecules at ordinary temperatures. It can be calculated from the molecular structure. The last column refers to the nanoemulsification capacity of the multiconnected micellar phase with respect to the respective fragrance, which is determined in Section 6.2.

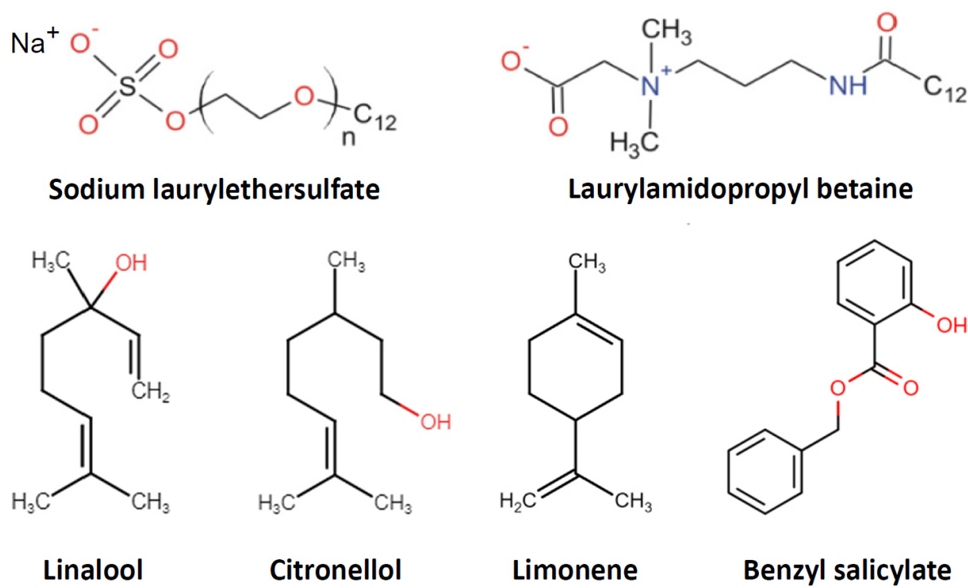

**Fig. 2.** Structural chemical formulas of the used surfactants and fragrances (oils).

**Table. 1** Properties of the fragrances used in the present article.

| Fragrance | log$P^{\dagger}$ | Water solubility, $S_w$ (mg/l)* | van der Waals volume (Å$^3$)$^{\dagger}$ | Nanoemulsification capacity, $\chi$ (mM/M)$^{\ddagger}$ |
|---|---|---|---|---|
| Linalool | 2.65 | 1589 | 174.9 | 1280 |
| Citronellol | 2.75 | 307 | 182.2 | 640 |
| Limonene | 3.22 | 13.5 | 154.7 | 1860 |
| Benzyl salicylate | 4.05 | 8.8 | 206.4 | 50 |

$^{\dagger}$ Calculated in Chemicalize [60].
*Data from Yalkowski et al. [61], Massaldi & King [62] and PubChem [63].
$^{\ddagger}$ Nanoemulsification capacity of the multiconnected micellar phase (Section 6.2).

*2.2. Methods*

*Preparation of the solutions.* First, the aqueous solution of the two basic surfactants, CAPB and SLES, was prepared. Next, the salt ($CaCl_2$, $ZnCl_2$ or $MgCl_2$) was added. The obtained solution was stirred for 1 hour at 60 °C. All measurements were carried out after equilibration of the solution for one night at 25 °C. The solutions were prepared with deionized water from the Millipore system (Milli- Q purification system, USA) with conductivity of 0.067 μS/cm.

*Optical observations* were carried out with a microscope AxioImager M2m (Zeiss, Germany) equipped with a fluorescence assay. Dyes were used to amplify the contrast between the water and micellar phases. The oil-soluble fluorescent dye BODIPY (Sigma Aldrich, 790389) colors the oily part of the bicontinuous phase yellowish-green. BODIPY is a fluorescent marker with excitation maximum at 502 nm and emission maximum at 510 nm (green light). To color the aqueous phase, we used the water-soluble dye methylene blue (Sigma Aldrich, M9140). In the solubilization/ nanoemulsification experiments, to color the oily phase (the fragrances), the red dye Sudan III (Sigma Aldrich, S4131) was used. Optical observations of the samples by cross-polarized light microscopy (CPLM) were carried out to verify whether any liquid crystalline mesophases have been formed. For the same purpose, the samples were studied also by small angle X-ray scattering (SAXS) with diffractometer D8 Advance ECO (Bruker, Billerica, Massachusetts, USA).

*Zeta potentials* of droplets from the multiconnected micellar phase were measured by a light scattering system Zetasizer Nano ZS (Malvern Instruments, UK).

*Rheological measurements* were carried out by a Bohlin Gemini apparatus (Malvern Instruments, UK) equipped with a CP60 flat-parallel plate. The shear rate was varied from 0.1 to 1000 s$^{-1}$. The temperature was controlled by a Peltier element; evaporation from the studied sample was prevented.

*Particle size distributions* in nanoemulsions were measured by two methods. First, laser particle sizer ANALYSETTE 22 NanoTec (Fritsch, Germany) with a measuring range of 0.01 – 2100 µm (laser diffraction method). Second, dynamic light scattering apparatus Malvern 4700C (Malvern Instruments, UK) – goniometric light scattering system for dynamic and static light scattering combined with a solid state laser (532 nm).

### 3. Effect of salts with divalent counterions

The experimental finding that the addition of Ca$^{2+}$ ions to a mixed solution of anionic and zwitterionic solution leads to the separation of isotropic micellar phase was serendipitous (observed when studying the effect of hard water on shampoo-type formulations). Next, we established that the phenomenon is more general and is observed also with other divalent counterions: Zn$^{2+}$ and Mg$^{2+}$. The effect is especially strong with Mg$^{2+}$ ions.

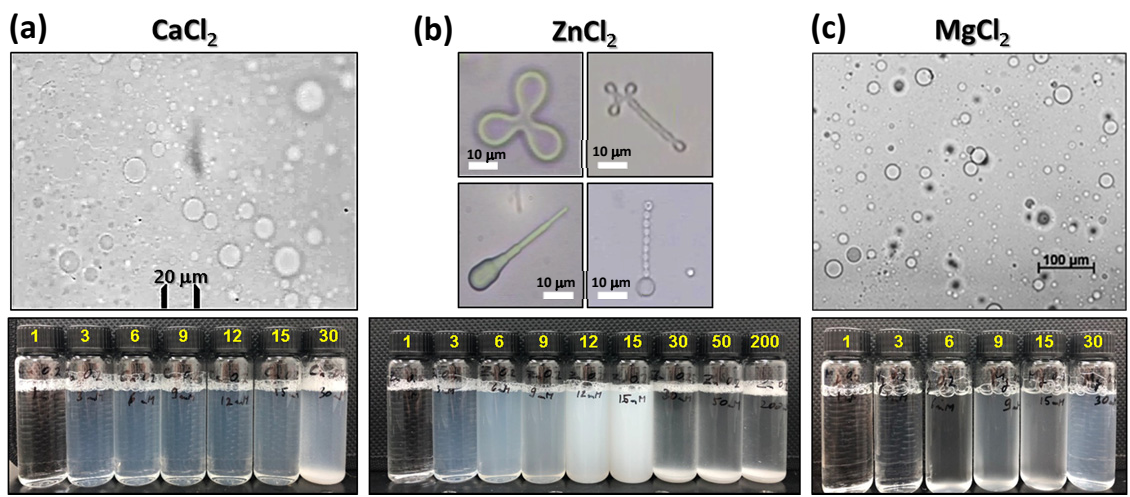

**Fig. 3.** Photographs illustrating the phase behavior of 0.2 wt% 1:1 w/w CAPB:SLES solutions with the rise of the concentration (mM) of added salt given on the caps of the vials: (a) CaCl$_2$; (b) ZnCl$_2$, and (c) MgCl$_2$. The photomicrographs (at the top) are taken for the solutions with 9 mM of the respective salt.

More specifically, we studied the phase behavior of CAPB + SLES solutions upon the addition of CaCl$_2$, ZnCl$_2$ and MgCl$_2$. The results are shown in Figs. 3 and 4. In Fig. 3, all solutions contain surfactant of total concentration $C_{tot}$ = 0.2 wt% 1:1 w/w CAPB:SLES plus added salt at different concentrations (mM) shown on the caps of the vials. The micrographs of the solution in Fig. 3 have been taken for the vials with 9 mM added salt.

In Fig. 4, the total surfactant concentration, $C_{tot}$, is varied at fixed weight ratio of the two surfactants (1:1 CAPB:SLES) and fixed 12 mM concentration of the added salt. The results are presented and discussed separately for each salt.

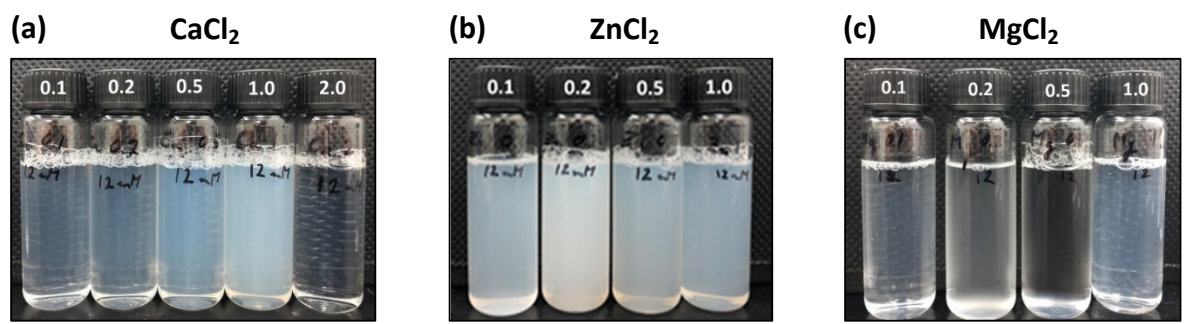

**Fig. 4.** Effect of the total surfactant concentration (wt%, given on the caps of the vials) on the solutions' phase behavior at a fixed surfactant composition, 1:1 w/w CAPB:SLES, and 12 mM (fixed) concentration of added salt: (a) $CaCl_2$; (b) $ZnCl_2$, and (c) $MgCl_2$.

*3.1. Effect of $CaCl_2$*

In Fig. 3a, for the surfactant solutions with $CaCl_2$ one can distinguish three types of phase behavior: A – clear solution; B – translucent (slightly turbid) solution, and C – precipitated (turbid) solution. Type A is observed with 1 mM added $CaCl_2$; this is a clear micellar solution ($C_{tot}$ is above the CMC). Type B is observed in the concentration range $3 \leq C_{salt} \leq 15$ mM ($C_{salt}$ denotes the $CaCl_2$ concentration). Type C is observed at $C_{salt}$ = 30 mM, where crystallites are formed in the solution; the smaller of them are dispersed in the bulk, whereas the bigger ones sediment at the bottom of the vial.

The microscopic observations show that the turbidity of *system type B* is due to the formation of drops with diameter on the order of 10 μm, as illustrated in the micrograph for 9 mM $CaCl_2$ in Fig. 3a. In cross-polarized light + lambda plate, these drops do not show any birefringence typical for ordered phases. Hence, the drops are not formed from liquid-crystalline mesophases like hexagonal, lamellar or spherulites (onions) [64-67]. The viscosity of the disperse systems of type B (drops in water) is close to that of water.

The drops are heavier than water and with time they sediment at the bottom of the vial. In the sediment, the droplets with $CaCl_2$ do not coalesce, so that they can be easily re-dispersed by shaking. A possible explanation is that the droplets are stabilized by electrostatic (double layer) repulsion; see Section 4.3 for the measured zeta-potentials. However, if subjected to centrifugation at an acceleration of 3000$g$ for 30 min, these drops coalesce and form a separate phase; see SI Appendix, Fig. S1 (SI = Supplementary Information). As demonstrated below, the sediment can be identified as a multiconnected micellar phase (saturated network).

In Fig. 4a, the $CaCl_2$ concentration is fixed, $C_{salt}$ = 12 mM, whereas the total surfactant concentration, $C_{tot}$, is increased from 0.1 to 2 wt% (1:1 w/w CAPB:SLES). For $0.1 \leq C_{tot} \leq 1$ wt%, one observes systems of type B with increasing turbidity. This can be explained with the increase of the volume fraction of the formed droplets with the rise of surfactant concentration. Finally, at $C_{tot}$ = 2 wt% the solution becomes clear (system type A) and its viscosity is markedly higher than that of water. This is indication for the presence of uniform micellar phase composed of entangled wormlike or branched micelles. The most probable interpretation is that at the highest surfactant concentration, $C_{tot}$ = 2 wt%, the amount of $CaCl_2$ is not enough for the formation of droplets from multiconnected micellar phase.

## 3.2. Effect of $ZnCl_2$

Fig. 3b illustrates the phase behavior of 0.2 wt% 1:1 w/w CAPB:SLES solutions upon the addition of $ZnCl_2$. With 1 mM $ZnCl_2$, a clear micellar solution is observed – system of type A. Systems of type B, translucent solutions with dispersed droplets, are observed in the concentration range $3 \leq C_{salt} \leq 9$ mM $ZnCl_2$. Milky-white precipitated system type C1 with dispersed submicron crystallites is observed at $12 \leq C_{salt} \leq 15$ mM, and precipitated system type C2 with sediment of crystallites is observed at $30 \leq C_{salt} \leq 200$ mM added $ZnCl_2$. The solutions with $ZnCl_2$ exhibit a stronger tendency to precipitation as compared with the solutions containing $CaCl_2$ and $MgCl_2$ (Fig. 3).

For the systems type B, in cross-polarized light + lambda plate the spherical drops with $ZnCl_2$ do not show birefringence typical for crystal phases. However, we observe also the presence of particles of non-spherical shape, like those shown in the micrograph in Fig. 3b (taken at 9 mM $ZnCl_2$), which show birefringence in polarized light. Hence, in the case of $ZnCl_2$ we observe coexistence of particles with liquid crystalline structure and particles from an isotropic phase (supposedly, multiconnected micellar phase).

In Fig. 4b, one sees that the turbidity does not monotonically depend on the total surfactant concentration, $C_{tot}$, at fixed (12 mM) $ZnCl_2$ concentration. Only the solution with $C_{tot}$ = 0.2 wt% is milky-white and precipitated. It seems that at $C_{tot}$ = 0.1 wt% the surfactant concentration is insufficient for precipitation, whereas at $C_{tot} \geq 0.5$ wt% the surfactant micelles have "solubilized" the crystallites.

*3.3. Effect of $MgCl_2$*

Fig. 3c illustrates the phase behavior of 0.2 wt% 1:1 w/w CAPB:SLES solutions upon the addition of $MgCl_2$. Systems of type A (clear micellar solutions) are observed for $C_{salt}$ = 1 and 3 mM $MgCl_2$. Systems of type B (translucent solutions with dispersed droplets), are observed in the concentration range $6 \leq C_{salt} \leq 15$ mM $MgCl_2$. The micrograph in Fig. 3c shows the droplets formed in a solution with 9 mM $MgCl_2$; some of them look darker because they are out of the focal plane of the microscope. A system of new type D is observed at $C_{salt}$ = 30 mM $MgCl_2$ – the solution is slightly more turbid but viscous (unlike the systems of type B). These properties indicate coexistence of precipitated crystallites with entangled wormlike or branched micelles.

The system of type B with $MgCl_2$ possesses a remarkable property. Upon sedimentation at the bottom of the vial, the drops spontaneously coalesce and form a separate phase. The bigger drops sediment and coalesce during the first 15 min. After 1 day at rest, practically all drops have merged into one large phase at the bottom. In Fig. 3c, the sedimented phase is not well visible because its refractive index is close to that of water. (In Section 4, the sedimented phase is visualized by the addition of dyes.) In Fig. 4c, all solutions correspond to type B, but some of them are clearer because of the sedimentation of a part of the drops.

Samples of the sedimented phase with $MgCl_2$ were subjected to observations in polarized light – no birefringence indicating liquid crystal mesophase was observed (just like in the case of drops with $CaCl_2$); see SI Appendix, Fig. S3. It is known that isotropic micellar phases and *cubic* liquid-crystal mesophases do not exhibit birefringence (colored patterns) in cross-polarized light with λ-plate [64,68,69].

Samples of the sedimented phase with $MgCl_2$ were subjected also to SAXS investigation. The shape of the SAXS spectrum (see, e.g., SI Appendix, Fig. S4) indicates that the sedimented phase is not a cubic phase, or any other liquid-crystal mesophase; see, e.g., Refs. [57,58,69-71]. Then the only possibility is it to be an isotropic saturated network from crosslinked wormlike surfactant micelles (Fig. 1), which form a separate phase. The existence of such phases was predicted and observed in preceding studies as a limiting case of branched micelles at high electrolyte concentrations [6-9,15]. The bicontinuous structure of the sedimented phase is confirmed by using lipophilic and hydrophilic dyes – see Section 4. Moreover, the position of the phase-separation domain in the viscosity salt curve (Section 5) is in full agreement with the theoretical predictions for the appearance of saturated network [11,12,30].

## 4. Visualization by dyes and zeta-potential measurements

### 4.1. Effects of salt concentration and CAPB/SLES ratio

Our subsequent investigations are focused on the systems with $MgCl_2$, because only in this case the separated surfactant-rich droplets spontaneously coalesce and form a separate phase, and the volume of this phase is the greatest at identical other conditions. To visualize its presence, the lipophilic dye BODIPY has been dissolved in the surfactant solution, where it is solubilized in the micelles. Upon phase-separation of the droplets of saturated micellar network and their sedimentation at the bottom of the vial, BODIPY is transferred in the sediment, which acquires yellow-green color – see Fig. 5a. The excess, non-solubilized dye represents reddish crystals that are seen at the bottom of some vials.

Fig. 5b shows the micellar phase separated from a solution of 0.5 wt% 1:1 (w/w) CAPB:SLES at six concentrations of $MgCl_2$, from 8 to 18 mM. One sees that the volume of the sedimented phase markedly decreases with the rise of $MgCl_2$ concentration. This can be due to the interplay of two effects. First, the increased electrolyte concentration suppresses the electrostatic swelling of the micellar network. Second, at higher salt concentrations water molecules can migrate from the multiconnected micellar phase to the bulk aqueous phase that leads to shrinkage of the micellar network (salting-out effect).

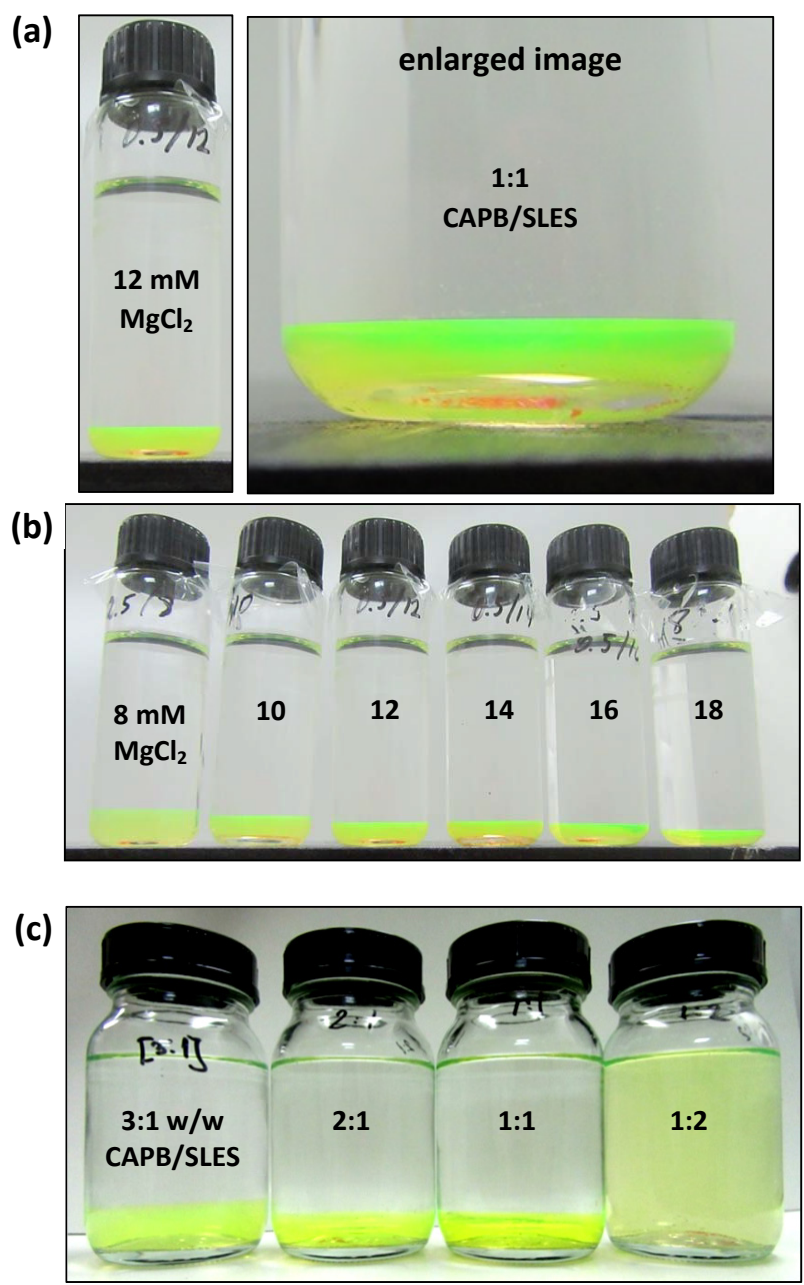

**Fig. 5.** Visualization of the sedimented multiconnected micellar phase with solubilized lipophilic dye BODIPY. (a) At surfactant concentration of 0.5 wt% 1:1 w/w CAPB:SLES with 12 mM $MgCl_2$. (b) The same surfactant solution at various $MgCl_2$ concentrations (mM) denoted on the vials. (c) Effect of surfactant composition (w/w, denoted on the vials) at fixed total surfactant concentration of 1 wt% and at fixed 30 mM $MgCl_2$.

In Fig. 5c, the total surfactant and salt concentrations are fixed ($C_{tot}$ = 1 wt% and $C_{salt}$ = 30 mM $MgCl_2$), whereas the composition of the surfactant mixture is varied from 3:1 to 1:2 (w/w) CAPB:SLES, that is from 25 to 67 % mass fraction of SLES relative to the total mass of surfactant. The photographs are taken one week after the preparation of the solutions.

Separation of micellar phase is observed in the range from 25 to 50 % SLES. In contrast, at 67 % SLES (1:2 CAPB:SLES) the solution is uniform (no phase separation) and more viscous. In this case, we are probably dealing with uniformly distributed entangled branched or wormlike micelles, rather than with saturated micellar network; see e.g. [9,11].

For the first three vials in Fig. 5c the volume of the separated micellar varies with the surfactant composition at fixed $C_{tot}$ = 1 wt% and $C_{salt}$ = 30 mM. This effect is probably related to the different micelle surface charge and counterion ($Mg^{2+}$) binding inside the network for different fractions of SLES in the surfactant blend.

*4.2. Experiments with hydrophilic and lipophilic dyes*

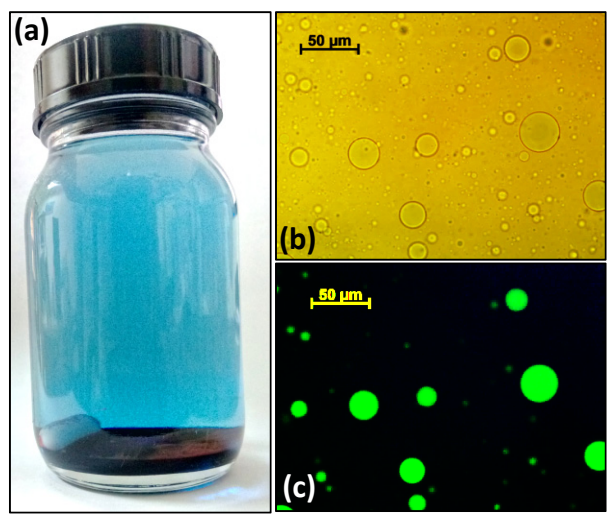

**Fig. 6.** Confirmation of the bicontinuous structure of the separated phase by the lipophilic dye BODIPY solubilized in the micellar phase and the hydrophilic dye methylene blue dissolved in the water phase; the solution contains 0.5 wt% 1:1 w/w CAPB:SLES and 16 mM $MgCl_2$. (a) Added methylene blue penetrates in the sedimented phase and colors it dark blue. (b) Drops from the sedimented phase dispersed in the supernatant (transmitted light). (c) The same drops glow green in fluorescence mode owing to the solubilized BODIPY.

If the separated phase is a saturated network of wormlike micelles (rather than compact drops of salted-out surfactant), it should contain continuous water phase. The latter is confirmed in an experiment with the water-soluble dye methylene blue – after its addition in the vial with the phase-separated solution, it penetrates in the sedimented phase, which acquires dark blue color (Fig. 6a). After dispersing the dark blue phase in the upper aqueous phase, in transmitted light we see that the dispersed droplets also have bluish color (Fig. 6b).

This time the intensity of the blue color in the drops is only slightly higher than that of the water background, because the thickness of the drops is much smaller than the thickness of the surrounding aqueous phase. If we cover the light source with 502 nm filter (the excitation wavelength of BODIPY), we see the same drops (as in Fig. 6b) shining in green color – Fig. 6c.

The fact that the sedimented surfactant-rich phase engulfs both the hydrophilic methylene blue (located in the aqueous phase) and the lipophilic BODIPY (solubilized in the micelles) confirms that we are dealing with a bicontinuous phase. This fact, combined with the lack of birefringence in polarized light and the shape of the SAXS diffraction spectrum (SI Appendix, Figs S3 and S4), as well as by the shape of the viscosity salt curve (Section 5) proves that the dispersed drops, and the bulk phase obtained after their coalescence, are composed of multiconnected micellar phase (saturated network); see Section 5.2 for a more detailed discussion.

It should be also noted that the dark background around the drops in Fig. 6c evidences that practically the whole amount of surfactant is concentrated in the multiconnected micellar phase, which should coexist with surfactant monomers in the aqueous phase. The fraction of smaller micelles (that could also solubilize BODIPY) in the aqueous phase turns out to be negligible.

The experiments illustrated in Fig. 6 were carried out in the presence of $MgCl_2$. Similar experiments with $CaCl_2$ and $ZnCl_2$ (instead of $MgCl_2$) lead to analogous results (see SI Appendix, Fig. S2). In other words, the drops separated in the solutions with $CaCl_2$ and $ZnCl_2$ also represent pieces of multiconnected micellar phase, despite the fact that they are more stable against coalescence and do not spontaneously merge into a large bicontinuous phase. However, such phase can be separated by centrifugation of the respective solution. The greatest yield of multiconnected micellar phase has been observed with $Mg^{2+}$. It seems that in the case of $Ca^{2+}$ and $Zn^{2+}$ a significant part of surfactant remains in the upper aqueous phase in the form of separate micelles. As already mentioned, in the case of $Zn^{2+}$ we observe also a population of non-spherical particles of liquid crystalline structure, which can be distinguished in polarized light, and which coexist with the spherical drops from the multiconnected micellar phase.

*4.3. Results from zeta-potential measurements*

To check whether the stability of the drops from the multiconnected micellar phase in the cases of $Ca^{2+}$ and $Zn^{2+}$, and the drop coalescence in the case of $Mg^{2+}$ is related to electrostatic double-layer interactions, we carried out zeta-potential measurements with such drops. For this goal, the sediment of multiconnected micellar phase was re-dispersed to μm-sized droplets in the aqueous phase above the sediment. The sedimentation of the obtained droplets was relatively slow, so that they were convenient for zeta-potential measurements. The composition of the used solutions was 0.5 wt% 1:1 w/w CAPB:SLES with added 16 mM $CaCl_2$, or $MgCl_2$, or $ZnCl_2$. The results (average from three runs) are shown in Table 2.

**Table 2**. Results from zeta-potential measurements with droplets from the multiconnected micellar phase separated in a solution of 0.5 wt% 1:1 w/w CAPB:SLES with 16 mM added salt.

| Salt | Zeta-potential (mV) | Conductivity (mS/cm) |
|---|---|---|
| $CaCl_2$ | −31.3 | 4.65 |
| $MgCl_2$ | −20.4 | 4.52 |
| $ZnCl_2$ | −17.7 | 4.61 |

In Table 2, the close values of conductivity for the three different salts indicate that the ionic strength in the aqueous medium is almost the same. Then, the difference between the values of the zeta-potentials could be due to different specific binding of the $Ca^{2+}$, $Mg^{2+}$ and $Zn^{2+}$ ions to the headgroups of SLES expressed on the surface of the drops from the multiconnected micellar phase. The negative sign of the zeta-potential indicates that the binding of divalent counterions does not invert the sign of the drop surface charge. In the case of $Ca^{2+}$, the magnitude of the zeta-potential is the highest, which is in agreement with the fact that the coalescence of drops is suppressed in the presence of $Ca^{2+}$. In the case of $Zn^{2+}$, the zeta potential is the lowest by magnitude, which indicates higher degree of counterion binding. The stability of the formed droplets against coalescence could be attributed to the short-range hydration repulsion [72] due to the bound strongly hydrated $Zn^{2+}$ ions [73]. In the case of $Mg^{2+}$, the drop coalescence can be a result of specific ability of the magnesium ions to interact with the surfactant headgroups, to bridge between neighboring drops and promote their merging.

## 5. Rheology of the multiconnected micellar phase

### 5.1. Salt curve and flow curves

In these experiments, the input surfactant concentration was fixed to 5 wt% 1:1 w/w CAPB:SLES. Fig. 7a shows the dependence of zero-shear viscosity, $\eta_0$, on the concentration of added $MgCl_2$. This dependence has the shape of a typical "salt curve" with a maximum [3−5]. The interpretation of the salt curve, which was established and confirmed in numerous works [3-5,8-13], is as follows. To the left from the maximum, $\eta_0$ increases with salt concentration because of the growth and entanglement of wormlike micelles. The maximum of $\eta_0$ marks the appearance of branched micelles. The degree of branching increased with the further growth of salt concentration. This is accompanied with a lowering of $\eta_0$ because of the mobility of the junctions in the branched micelles [15]. At sufficiently high salt concentrations, the connection of all branched micelles leads to phase separation of saturated micellar network [8-12]. Despite the relatively high $MgCl_2$ concentrations (up to 90 mM) all solutions are clear (no precipitation), which can be explained with the binding of $Mg^{2+}$ ions to the micellar aggregates.

In Fig. 7a, the viscosity $\eta_0$ of the phase-separated saturated micellar network (the four rightmost points) is compared with the values of $\eta_0$ for the homogeneous phase before the phase separation (all other experimental points). For the multiconnected micellar phase, $\eta_0$ turns out to be constant (independent of the input $MgCl_2$ concentration) and relative low, as expected [15]. The separation of isotropic surfactant phase at the highest studied salt concentrations (Fig. 7a) is in full agreement with the physical interpretation of the salt curves [3-5,8-13], and is an important argument in favor of the identification of the separated phase as a saturated micellar network.

Fig. 7b shows flow curves, which represent plots of the apparent viscosity $\eta$ (the ratio of the shear stress and shear rate) vs. the shear rate $\dot{\gamma}$ obtained by the rotational rheometer in shear-rate-ramp regime. The four experimental curves are measured with the phase-separated saturated micellar network and correspond to the four rightmost points in Fig. 7a.

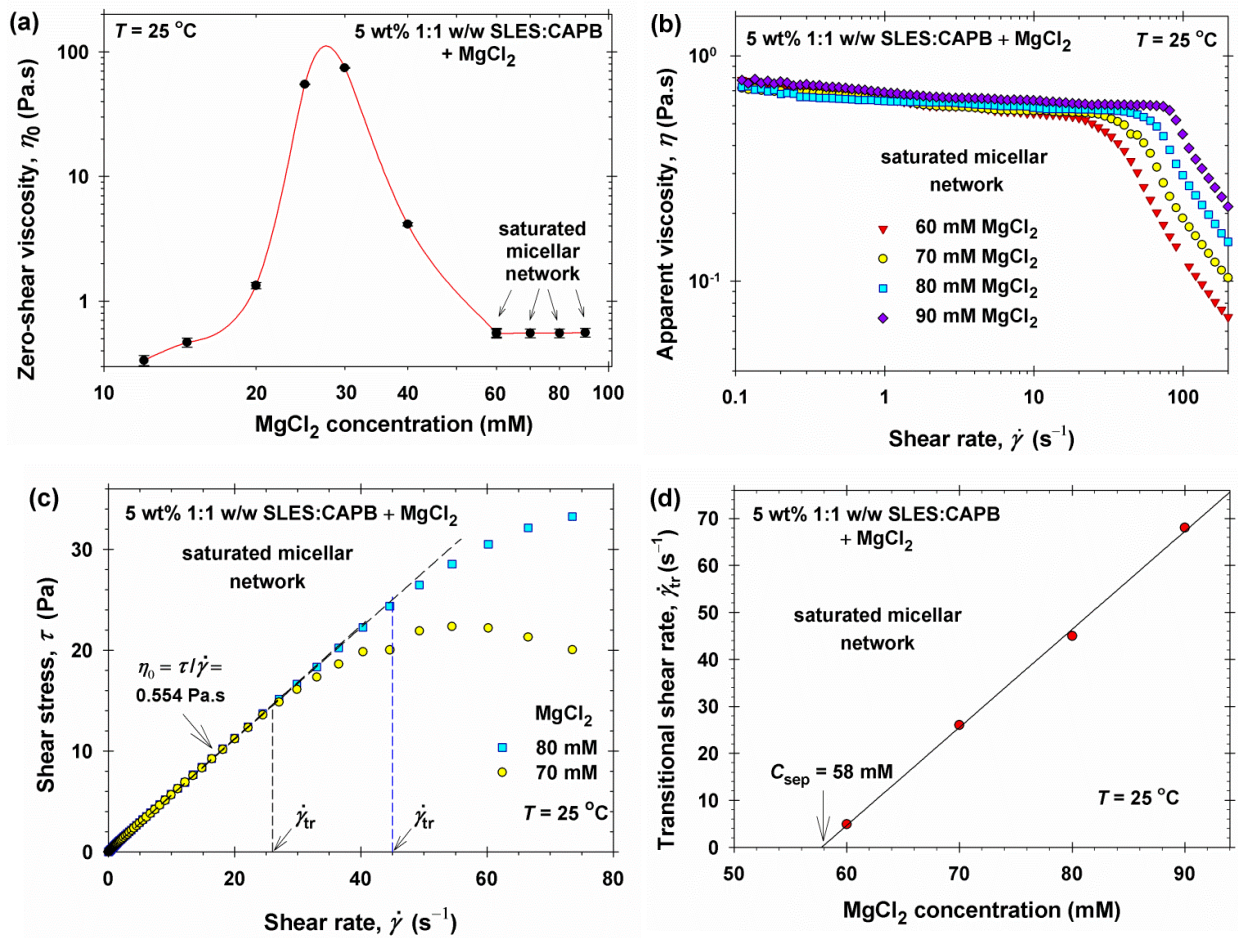

**Fig. 7.** Rheological properties of the micellar solutions of 5 wt% 1:1 w/w CAPB:SLES at various MgCl$_2$ concentrations. (a) "Salt curve": zero-shear viscosity, $\eta_0$, vs. MgCl$_2$ concentration; the rightmost four experimental points represent $\eta_0$ of the phase-separated saturated micellar network; the line is a guide to the eye. (b) Apparent viscosity, $\eta$, vs. the shear rate, $\dot{\gamma}$, for the saturated micellar network at four different MgCl$_2$ concentrations. (c) Shear stress $\tau$ vs. shear rate $\dot{\gamma}$ – determination of $\eta_0$ from the slope at $\dot{\gamma} \to 0$ and determination of the transitional shear rate, $\dot{\gamma}_{tr}$, that characterizes the transition from Newtonian to shear-thinning behavior. (d) Plot of $\dot{\gamma}_{tr}$ vs. the MgCl$_2$ concentration; $C_{sep}$ is the first MgCl$_2$ concentration, at which phase separation of saturated micellar network appears. In Figs. 7c and d the lines are linear regressions.

In Fig. 7b, one sees that at the lower shear rates, the apparent viscosity is almost constant, $\eta \approx \eta_0$, i.e. a quasi-Newtonian behavior is observed. At higher shear rates, the viscosity decreases with the rise of $\dot{\gamma}$, i.e. pseudoplastic behavior (shear thinning) is observed. Similarly shaped flow curves have been obtained with other concentrated surfactant solutions containing wormlike and branched micelles; see e.g. Refs. [5,23,25]. Newtonian behavior in a wide interval of shear rates was registered in Ref. [15] for saturated micellar networks.

To accurate determine the transitional shear rate, $\dot{\gamma}_{tr}$, between the regions with Newtonian and pseudoplastic behavior, we plotted the shear stress, $\tau$, vs. the shear rate $\dot{\gamma}$; an example is given in Fig. 7c. The onset of deviation of the experimental $\tau$-vs.-$\dot{\gamma}$ dependence from the straight line determines $\dot{\gamma}_{tr}$. In addition, the slope of the straight line determines $\eta_0$.

In Fig. 7d, the obtained values of $\dot{\gamma}_{tr}$ are plotted vs. the $MgCl_2$ concentration. The data are fitted with straight line, which intersects the horizontal axis at 58 mM $MgCl_2$. Experimentally, at concentrations $\leq 58$ mM $MgCl_2$ we do not observe the separation of droplets from the multiconnected micellar phase and their sedimentation at the bottom of the vessel. In other words, we could conclude that for a solution of input surfactant concentration 5 wt% 1:1 w/w CAPB:SLES, the saturated micellar network appears at a phase-separation concentration $C_{sep} \approx 58$ mM $MgCl_2$. Additional investigations are needed to determine systematically the dependence of $C_{sep}$ on the surfactant concentration and composition, which is out of the scope of the present study. The use of plots like Fig. 7c represents a method for achievement of this goal.

The observation of shear thinning (Fig. 7b) indicates that the shearing deformation produces structural changes in the multiconnected micellar phase. Fig. S5 in SI Appendix shows the variation of viscosity during three consecutive shear-rate ramps with the same sample. The period of rest between the first and second run is 60 s, whereas that between the second and third run is 30 s. The similar shapes of the three viscosity curves obtained in the three consecutive runs indicate that the initial structure of the disturbed multiconnected micellar phase is quickly and spontaneously rebuilt.

For all regions of the salt curve (Fig. 7a), rheological data in oscillatory regime for the storage and loss moduli ($G'$ and $G''$) will be published in a subsequent paper.

## 5.2. Discussion

From the viewpoint of rheology (Fig. 7a), the observed phase separation is in full agreement with the well-established consequence of transformations with the linear ionic surfactant aggregates with the rise of salt concentration, viz. wormlike micelles → branched micelles → saturated network. From a general viewpoint, there are only two possible explanations of the phase separation at high salt concentrations [9,11,12,30]:

(1) Precipitation (salting out) of surfactant; the precipitates are either solid or liquid crystals, which can be detected by CPLM and SAXS experiments [69].

(2) Separation of saturated network from wormlike micelles [8-12], which was initially termed L1/L1 phase separation [7].

It is important to note that the saturated network is the only kind of micellar (isotropic) structure, which phase-separates, as predicted theoretically [8], and confirmed in the subsequent studies in this field; see e.g. Refs. [6-15]. All other micellar structures do not phase-separate, but occupy the whole volume of the aqueous solution.

Our CPLM experiments (plus SAXS for the cubic phases) showed that the observed phase separation is not precipitation (salting out) of surfactant. Hence, only the second possibility remains, viz. the separated phase to be saturated network of wormlike micelles.

From a logical viewpoint, the CPLM experiments (plus SAXS for the cubic phases) are sufficient to conclude that we are dealing with saturated micellar network. Our additional experiments, viz. (i) staining with dyes (Fig. 6) and (ii) rheological salt curve (Fig. 7a), have auxiliary character – they only support and confirm the conclusions based on the CPLM+SAXS experiments. Other supporting experiments could be carried out with diffusion NMR and cryo-TEM, which could be subjects of future studies.

## 6. Nanoemulsification of oils by the multiconnected micellar phase

*6.1. Wormlike micelles vs. multiconnected micellar phase*

Illustrative results are shown in Fig. 8a and b, where the left bottle (with almost clear aqueous phase) represents a solution of 5 wt% 1:1 w/w CAPB:SLES (≈145 mM total surfactant) without $MgCl_2$, where short wormlike micelles are formed [26]. The right bottle has been filled with multiconnected micellar phase, which represents the sediment from a solution of 1 wt% 1:1 w/w CAPB:SLES plus 30 mM $MgCl_2$. From the volume of the sedimented multiconnected micellar phase relative to the total volume of the solution (Fig. 5c), one could estimate that the left and right bottles contain solutions of approximately equal surfactant concentrations, but with different initial micellar structures: wormlike micelles vs. saturated network. The viscosities of the two initial phases are close to each other, 0.3 – 0.4 Pa·s.

To visualize the location of the oil phase, it was colored red by a small amount of the oil-soluble dye Sudan III. In Figs. 8a and b, respectively, 1 wt% and 3 wt% limonene was added over the aqueous phase. Then, the solutions were subjected to mild agitation by magnetic stirrer for an hour at room temperature. The photographs taken afterwards show that the solution with wormlike micelles in the left bottles (Figs 8a and b) has solubilized only a tiny fraction of limonene, as evidenced by the pale pink color of the aqueous phase, whereas the main part of limonene is separated as a top layer of oil. In contrast, in the right bottles the whole oil phase has been absorbed by the surfactant solution, which has acquired intense red color and has become turbid. The turbidity is due to the presence of 100 – 200 nm droplets, as seen in Fig 8c. Hence, the interaction of the multiconnected micellar phase with the oil has led to the formation of oil-in-water nanoemulsion.

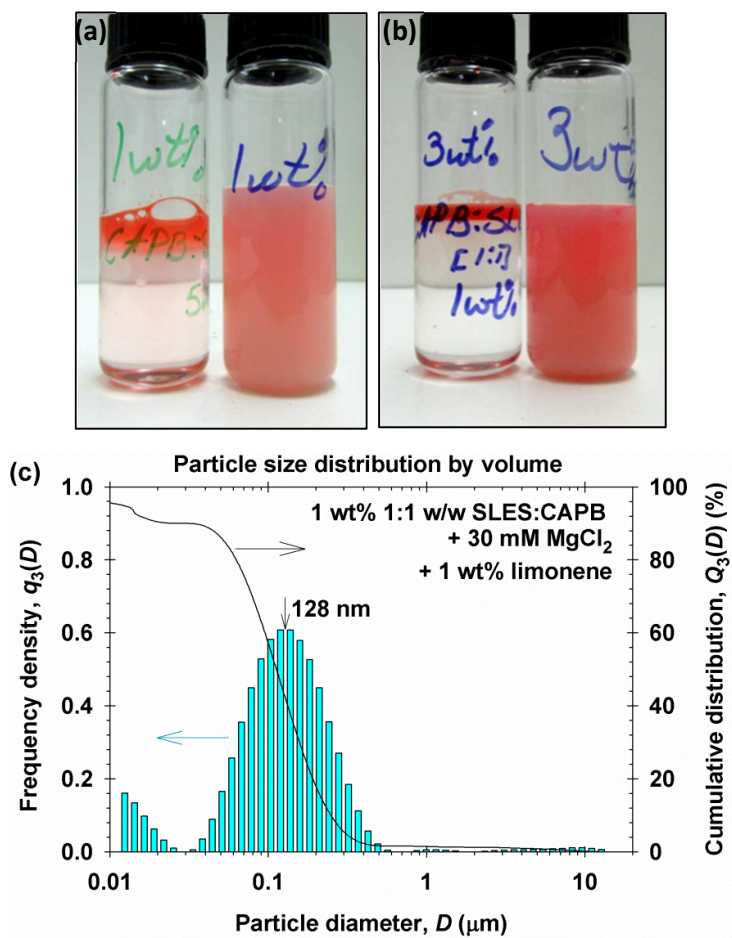

**Fig. 8.** Interaction of limonene with micellar solutions containing 1:1 w/w CAPB:SLES. (a,b) The left bottle contains 5 wt% micellar solution without $MgCl_2$ (no formation of saturated network), whereas the right bottle contains multiconnected micellar phase sedimented from 1 wt% surfactant solution + 30 mM added $MgCl_2$; (a) with 1 wt% added limonene; (b) with 3 wt% added limonene. (c) Size distribution of the limonene droplets in the right bottle in Fig. 8a.

To obtain the graph in Fig. 8c, the laser particle sizer was used. For this goal, the turbid dispersion in the right-hand-side bottle in Fig. 8a was diluted 150 times with pure water. The quantity $q_3(D)dD$ represents the volume fraction of particles (droplets) with diameters in the interval from $D$ to $D+dD$; $q_3(D)$ is the particle size frequency distribution. Correspondingly, the quantity

$$Q_3(D) = \int_D^\infty q_3(\hat{D}) d\hat{D} \tag{1}$$

is the cumulative particle size distribution; $\hat{D}$ is a variable of integration. Fig. 8c shows that the droplets in the nanoemulsion have lognormal size distribution, the most probable drop diameter being 128 nm. The small peak at $D < 30$ nm probably represents spheroidal and/or short rodlike micelles.

The dilution of the sample during the particle sizing by laser diffraction (Fig. 8c) has certainly affected the size of the surfactant self-assemblies present in the solution. Note, however, that the target of this measurement is the size of the produced emulsion drops, rather than the size of the micelles. The emulsion drops are expected to be stable, despite the dilution, because the surfactant concentration remains above the CMC. To confirm that, we carried out independent drop-sizing experiments by DLS. The result, which is shown in SI Appendix Fig. S7, is in very good agreement with that in Fig. 8c. We can conclude that the size distribution of the nanoemulsion drops has been correctly determined.

*6.2. Nanoemulsification capacity of the multiconnected micellar network*

By the addition of increasing amounts of oil (limonene, linalool, etc.) to the separated multiconnected micellar phase, we found that above a certain amount of oil it cannot be emulsified by the surfactant solution, but forms lenses floating above the aqueous phase (or sediment from the heavier benzyl salicylate at the bottom). Hence, the saturated micellar network has a certain nanoemulsification capacity. In analogy with the micellar solubilization capacity [74], the nanoemulsification capacity can be defined by the expression:

$$\chi = \frac{C_{sat} - S_w}{C_{tot} - CMC} \approx \frac{C_{sat}}{C_{tot}} \tag{2}$$

Here, $C_{sat}$ is the saturation concentration of oil in the aqueous surfactant solution; $S_w$ is the solubility of the respective oil in pure water, and $C_{tot}$ is the total surfactant concentration. All concentrations in Eq. (2) are expressed in mol/l relative to the total volume of the liquid, so that $\chi$ is a dimensionless quantity. We have used also the circumstance that in our case $C_{sat} \gg S_w$ and $C_{tot} \gg$ CMC.

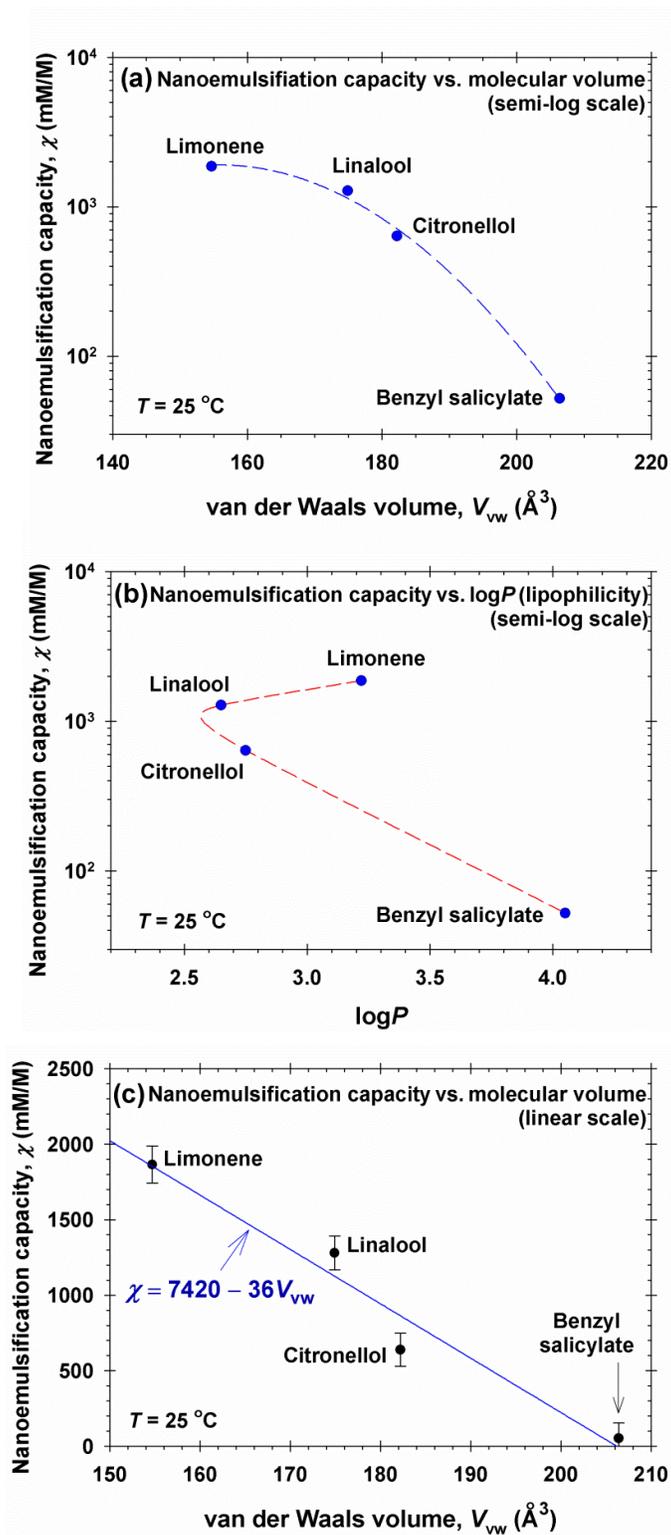

**Fig. 9.** Nanoemulsification capacity, $\chi$, of a bicontinuous phase containing 1 wt% 1:1 w/w CAPB:SLES + 30 mM $MgCl_2$: (a) $\chi$ vs. the van der Waals volume, $V_{vw}$, of the oil molecule in semi-log scale; the line is a guide to the eye; (b) $\chi$ vs. log$P$; the dashed line is a spline that connects the points in the same order as in Fig. 9a; (c) $\chi$ vs. $V_{vw}$ in linear scale; the line is a linear regression.

In this series of experiments, we produced a greater amount of the multiconnected micellar phase, which was separated as sediment from an initial aqueous solution of 1 wt% 1:1 w/w CAPB:SLES with 30 mM $MgCl_2$. To determine $C_{tot}$ in the sedimented phase, a portion of it was dried to constant weight. The amount of surfactant was estimated by subtracting the amount of $MgCl_2$, assuming one Mg atom per two SLES molecules in the dried substance. Then, knowing the contents of the active, the sedimented phase was diluted by pure water to $C_{tot}$ = 1 wt% 1:1 w/w CAPB:SLES, which is the real surfactant concentration in the saturated network.

Different portions of colored (with Sudan III) oil were added by a micropipette to a series of vials filled with identical volumes of the multiconnected micellar phase thus prepared. Next, the content of the vials was agitated by magnetic stirrer for 1 hour, and then the vials were kept at rest for 3 weeks. $C_{sat}$ was determined as the mean arithmetic between the concentrations of oil in the first vial with separated oil phase and in the last vial without separated oil phase; see, e.g., the vials with citronellol in SI Appendix, Fig. S3.

Next, the nanoemulsification capacity of the multiconnected micellar phase, $\chi$, has been calculated using Eq. (2). The results for $\chi$ are shown in Table 1 and plotted in Figs. 9a and b as functions, respectively, of the van der Waals volume of the oil molecule, $V_{vw}$, and $\log P$ for this oil; see Table 1. One sees that $\chi$ monotonically decreases with $V_{wv}$, but $\chi$ does not correlate with $\log P$. If the nanoemulsification capacity was governed by the lipophilicity of the oil phase, then one could expect that $\chi$ should grow with the rise of $\log P$; however, such tendency is observed only for linalool and limonene (Fig. 9b). The correlation between $\chi$ and $V_{wv}$ is demonstrated also in Fig. 9c, where the same data (as in Fig. 9a) are plotted in linear scale. One sees that in first approximation the dependence of $\chi$ on $V_{wv}$ is linear. This dependence indicates that the mechanism of production of nanoemulsions, as a result of interaction of the saturated micellar network with the oil, is governed by steric (in addition to the hydrophobic) interactions.

Fig. 9c indicates that for $V_{vw} > 207$ Å$^3$ oil separation always takes place. Note, however, that nanoemulsions can be formed even for oil concentrations > $C_{sat}$, for which the turbid o/w nanoemulsion coexists with separated oil phase. As an example, see the data for benzyl salicylate in SI Appendix, Fig. S3, where the non-emulsified oil (colored red with Sudan III at the bottom of vials) coexists with a multiconnected micellar phase (the clear phase) and with a nanoemulsion (the turbid phase). Such phase behavior resembles that of Winsor III systems [75,76], but its investigation is out of the scope of the present study. We observed nanoemulsification by multiconnected micellar phase also with linear alkanes of not too long paraffin chain, viz. octane and decane.

*6.3. Discussion on nanoemulsification*

The solubilization of fragrances in the micelles of sodium dodecyl sulfate (SDS) was investigated by Fan et al. [77]. These authors concluded that the perfumes with a linear chain structure or an aromatic group can penetrate into the palisade layer and closely pack with the surfactant molecules in spherical or slightly elongated micelles (97.5 mM SDS, no salt).

In our case, the fragrances interact with a network of multiconnected (crosslinked) wormlike micelles (Fig. 1). The fact that (unlike the saturated network) the wormlike micelles in contact with oil do not produce nanoemulsions (see Fig. 8), implies that the junctions (nodes) in the micellar network play a central role in nanoemulsification. Indeed, the junctions are the structural units that make the difference between wormlike micelles and saturated network. We could hypothesize that the interaction of the oil molecules with the micellar network includes solubilization of oil inside the junctions, which swell and detach as separate nanoemulsion drops. At that, the saturated micellar network is destroyed. This process deserves to be investigated in more details in the future.

Whatever the mechanism of nanoemulsification by the multiconnected micellar phase could be, it is much less energy consuming in comparison with the methods that employ mechanical drop breakup, like the high pressure homogenizers. In the case of multiconnected micellar phase, only a gentle mechanical stirring has been used to mix the two phases and to bring the oil in contact with the micellar network.

**7. Conclusions**

Although the phase separation of a saturated micellar network as a result of interconnection of branched micelles was predicted long ago [7,8], the observation of such phase separation was reported in only few studies as an element of experimental phase diagrams and/or as multiconnected domains observed by cryo-TEM [6,15,40,48,49]. To the best of our knowledge, the present article represents the second study, after Ref. [15], which is especially dedicated to the properties of multiconnected micellar phases, on the basis of production of macroscopic amounts of such phases, which were next subjected to investigation.

Phase separation of saturated micellar network was established in mixed solutions of the anionic surfactant SLES and the zwitterionic CAPB in the presence of divalent counterions, $Ca^{2+}$, $Zn^{2+}$ and $Mg^{2+}$, at concentrations below 0.1 M and at room temperature, 25 °C (Sections 3 and 4). It should be noted, that in the preceding study with cationic surfactant systems [6], the salt concentrations were much higher, 3–5 M, and the temperature range was 30–90 °C. In our case, the network formation could be attributed to binding of the divalent counterions to surfactant headgroups at the micellar surface, which suppresses the electrostatic repulsion and induces spontaneous packing parameter that favors the formation of intermicellar junctions [78]. For the studied systems, the saturated network appears in the form of drops (Figs. 3 and 4), which are heavier than water and sediment at the bottom of the vessel. In the case of $Ca^{2+}$ and $Zn^{2+}$, these drops are relatively stable and do not spontaneously coalesce in the sediment. In the case of $Zn^{2+}$, they coexist with particles from a liquid-crystal mesophase. In the case of $Mg^{2+}$, only drops from isotropic phase are formed and spontaneously merge into a separate micellar phase (Fig. 5) – a supergiant multiconnected micelle.

The structure of the separated phase is established by combining four experimental methods: (i) cross-polarized light microscopy (CPLM); (ii) small-angle X-ray scattering (SAXS); see SI Appendix Figs. S3 and S4; (iii) staining with dyes (Fig. 6), and rheometry – salt curve (Fig. 7a). From a logical viewpoint, the results from the first two methods, CPLM and SAXS, are sufficient to identify the separated phase as saturated micellar network; see Section 5.2 for details. The experiments by the other two methods (staining with dyes and rheometry), in spite of having auxiliary character, convincingly confirm the identification of the separated phase.

The rheological study of the multiconnected micellar phase in shear-rate-ramp regime shows quasi-Newtonian and shear-thinning regions in the flow curves (Fig. 7b). The plot of the transitional shear rate (Fig. 7c) versus the Mg concentration can be used as a tool for determining the onset of saturated network formation upon the addition of salt (Fig. 7d). After being subjected to intensive shearing, the multiconnected micellar phase quickly rebuilds its structure.

The addition of oils with relatively small molecules, like the fragrances limonene, linalool, citronellol and benzyl salicylate, leads to the formation of oil-in-water nanoemulsions (Fig 8) with a minimal energy input for mixing of the aqueous and oily phases. The nanoemulsification capacity of the multiconnected micellar phase correlates with the size of the oil molecule, but does not correlate with its lipophilicity characterized by log$P$ (Fig. 9). The experiments indicate that above a certain size of the oil molecule, complete nanoemulsification of the added oil is impossible (Fig. 9c). This result implies a possible role for the network junctions in the nanoemulsification process (see Section 6.3).

Future development of the present study could include the following research directions:

(i) Confirmation of the saturated network by cryogenic transmission electron microscopy (cryo-TEM).

(ii) Separation of bulk multiconnected micellar phases with other counterions (not only $Mg^{2+}$), and with other surfactants, and investigation of their properties.

(iii) Systematic investigation of the rheological properties of multiconnected micellar phases in different dynamic regimes and for different solution compositions and comparison with the theoretical predictions [8-12]; detection and characterization of the onset of network formation upon the rise of salt concentration.

(iv) Systematic study on nanoemulsification by multiconnected micellar phase – mechanism, factors affecting the drop size; appearance of microemulsion phases; rheology, etc.

The multiconnected micellar phases could find applications, at least, in three fields:

(i) Their ability to be dispersed to small droplets, which can solubilize target molecules and then coalesce again, can be used in some extraction and separation processes.

(ii) Stable droplets of saturated micellar network, like those with $Ca^{2+}$, can be used as carriers of various drugs, actives and fragrances, like the cubosomes and hexosomes [57-59].

(iii) The nanoemulsification of oily substances by multiconnected micellar phase can find applications for the production (at low energy cost) of nanoemulsions with their numerous applications [50-56].

**Acknowledgments**

The authors gratefully acknowledge the support from the National Science Fund of Bulgaria, Grant No. DN 09/8/2016 and from the Operational Programme ''Science and Education for Smart Growth", Bulgaria, project No. BG05M2OP001-1.002-0023. They thank Dr. Krastanka Marinova for her contribution to the management of this project; Dr. Filip Ublekov for the SAXS measurements, and Ms. Theodora Stancheva for a part of the rheological measurements.

**Appendix A. Supplementary information**

Supplementary material related to this article can be found, in the online version, at https://doi.org/10…..

# Supplementary Information
## for the article

# Phase separation of saturated micellar network and its potential applications for nanoemulsification

Mihail T. Georgiev, Lyuba A. Aleksova, Peter A. Kralchevsky*, Krassimir D. Danov

*Corresponding author: Peter A. Kralchevsky
 Email: pk@lcpe.uni-sofia.bg

# APPENDIX

(List of the references cited here is given at the end of this Appendix)

$C_{tot}$ = 0.5 wt% 1:1 w/w CAPB:SLES-1EO + 16 mM salt:

+ CaCl$_2$  +  ZnCl$_2$  +  MgCl$_2$

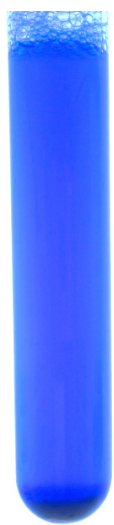
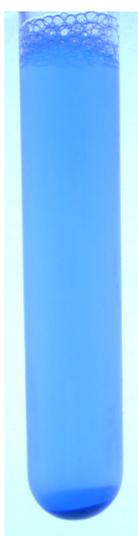
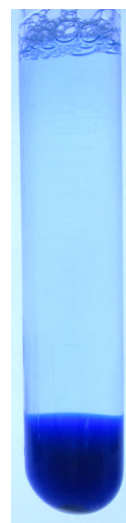

**Fig. S1.** Separation of the multiconnected micellar phase by centrifugation from a solution of total surfactant concentration $C_{tot}$ = 0.5 wt% 1:1 w/w CAPB:SLES-1EO + 16 mM salt: CaCl$_2$, ZnCl$_2$ and MgCl$_2$. The solutions have been subjected to centrifugation at 3000$g$ for 30 min. All solutions contain equal amounts of methylene blue (a water soluble dye), which is distributed between the sediment and supernatant. The photographs show different volumes of sediment in the presence of Ca$^{2+}$, Zn$^{2+}$ and Mg$^{2+}$ ions. The greatest yield of multiconnected micellar phase (the sediment) is observed with MgCl$_2$, for which the saturated network can be separated also by common gravitational sedimentation; the centrifugation only accelerates this process.

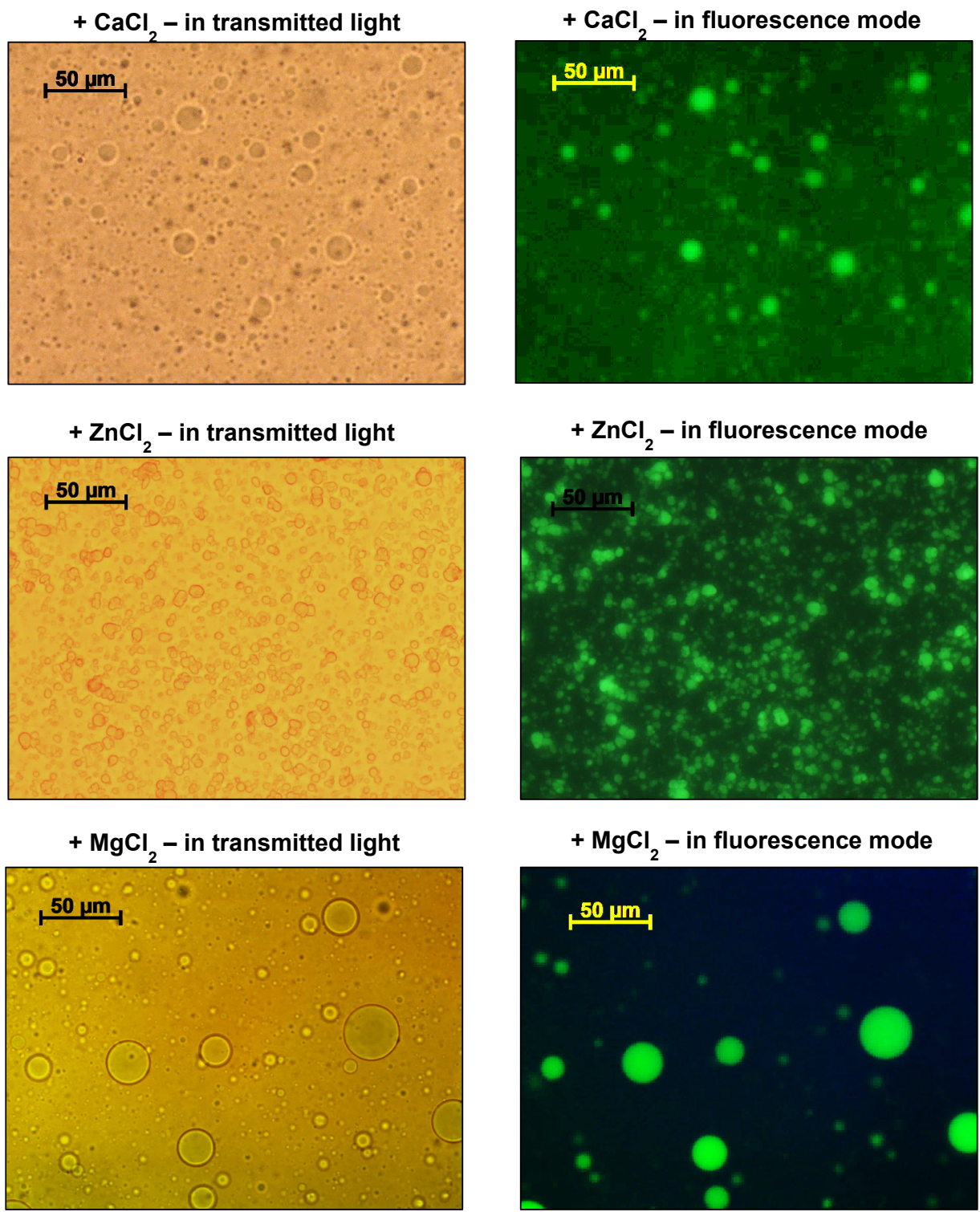

**Fig. S2.** Drops from the sedimented phase dispersed in the supernatant: Photomicrographs in transmitted light (left) and in fluorescence mode (right). In the latter case the same drops glow green owing to the solubilized lipophilic dye BODIPY. All solutions contain also methylene blue (Fig. S1).

(a)

(b)

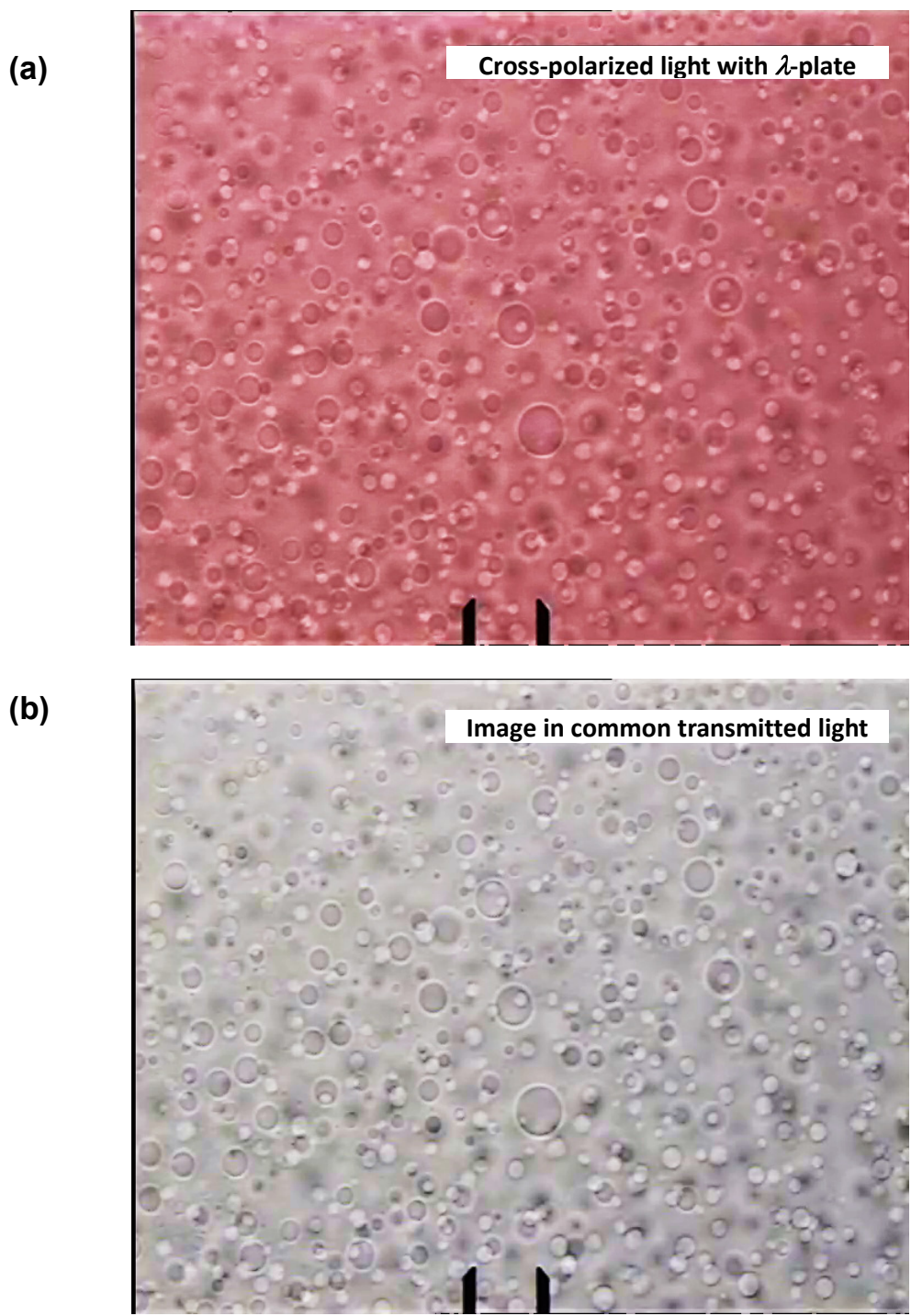

**Fig. S3.** Micrographs of drops from the sedimented phase dispersed in the supernatant; the total surfactant concentration is $C_{tot}$ = 1 wt% 1:1 w/w CAPB:SLES-1EO + 30 mM $MgCl_2$. (a) In cross-polarized light with λ-plate, the drops do not show any birefringence colors, which would indicate the formation of liquid-crystal mesophase; the observed uniform magenta color is typical for isotropic phases. (b) Control micrograph – the same drops in common transmitted light. The reference caliper shows 50 μm.

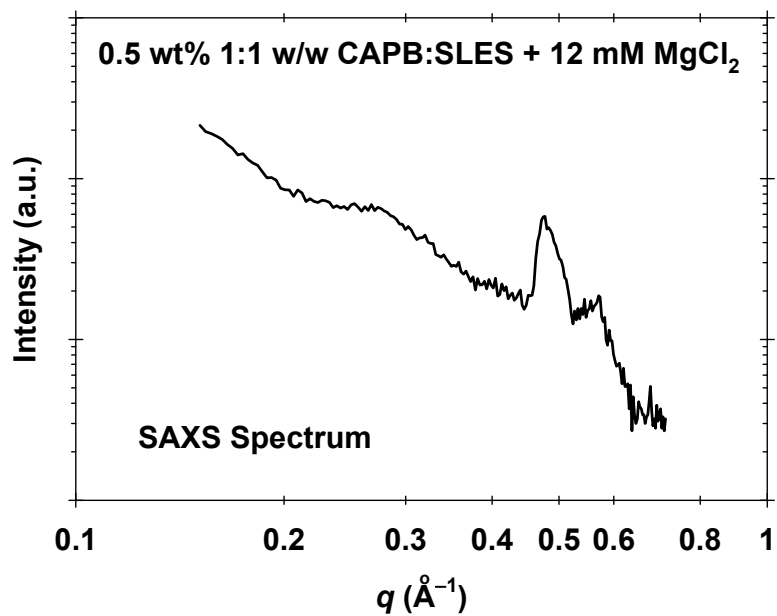

**Fig. S4.** SAXS spectrum from the surfactant-rich phase sedimented at the bottom of a solution of total concentration 0.5 wt% 1:1 w/w CAPB:SLES + 12 mM $MgCl_2$ – this is the system shown in the photos in Fig. 5a in the main text. The spectrum does not correspond to any ordered liquid-crystal mesophase, in particular – to cubic phase; see e.g. Refs. [1-5]. See comments on the SAXS spectrum on p. 6 below.

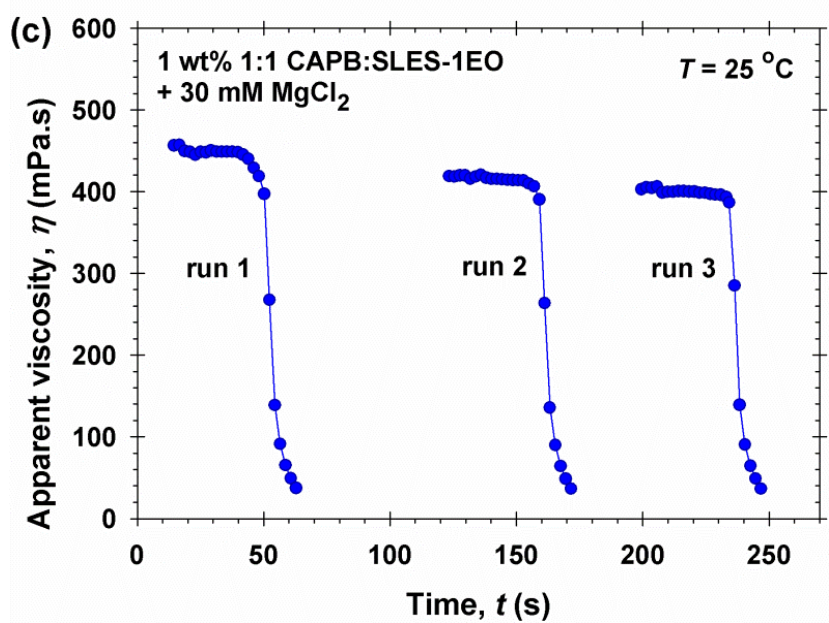

**Fig. S5.** Rheological properties of the multiconnected micellar phase separated from a solution of 1 wt% 1:1 CAPB:SLES in the presence of 30 mM $MgCl_2$: Plot of $\eta$ vs. time for three consecutive shear-rate-ramp runs with the same multiconnected micellar phase demonstrating the rebuilding of the micellar structure after intensive shearing.

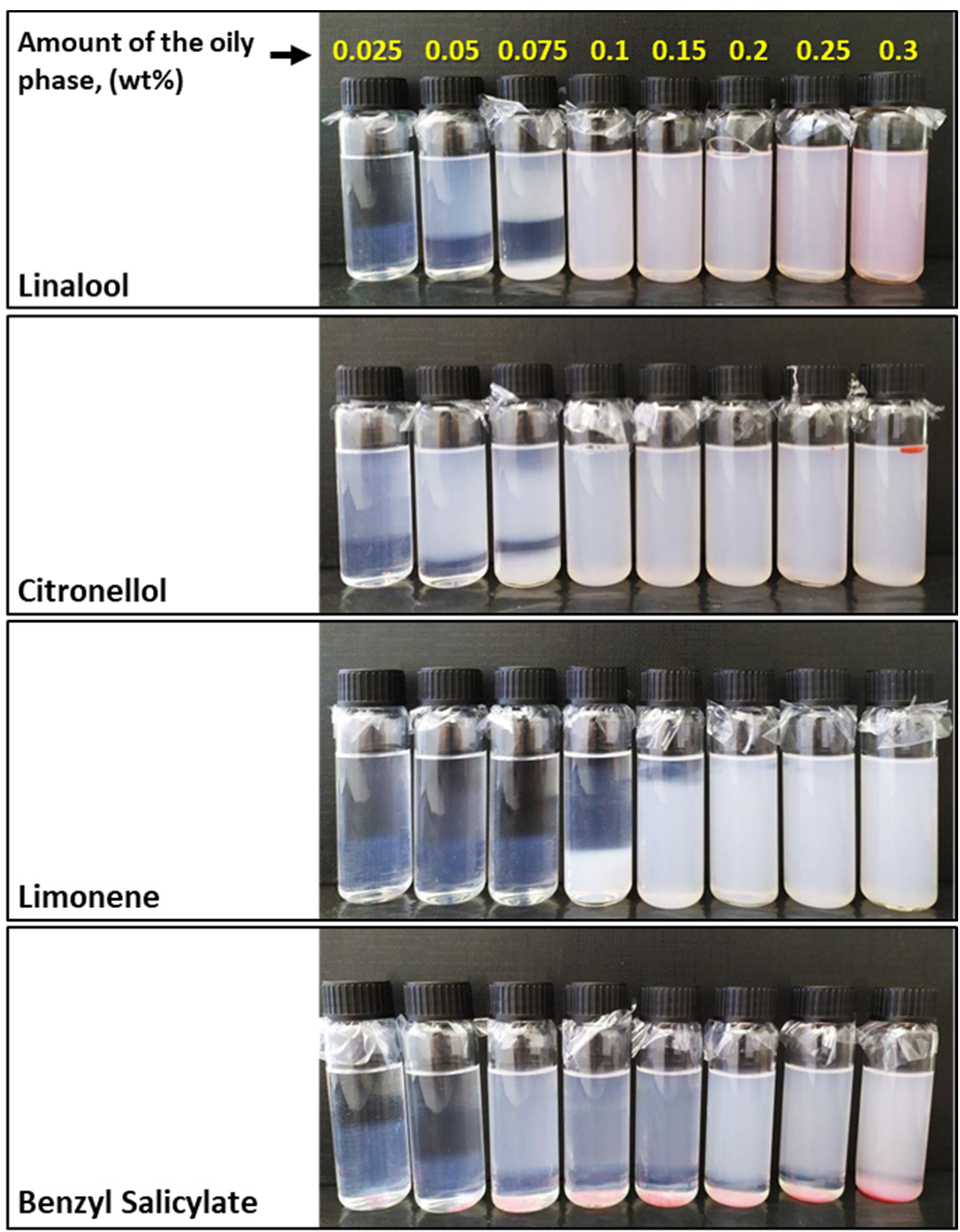

**Fig. S6.** Interaction of oil (fragrance) with a multiconnected micellar phase, which contains 1 wt% 1:1 w/w CAPB:SLES + 30 mM $MgCl_2$. The amount of added oil (in wt% relative to the total liquid phase) is denoted in the top line and refers to all investigated oils in the same column. Initially, the multiconnected micellar phase was poured in the vial. Next, the respective amount of oil, colored red with the dye Sudan III, was added and the content of the vial was agitated by magnetic stirrer for 1 h. After that, the vials were kept at rest for 3 weeks and, then, the photographs shown above were taken. One sees that at the lower oil concentrations, we have phase separation of multiconnected micellar phase (clear) and nanoemulsion (translucent or turbid); the latter is heavier or lighter depending on the oil type and concentration. At the higher oil concentrations, the nanoemulsion occupies the whole vial. In the case of benzyl salicylate, which is heavier than water, we observe also three-phase equilibria, at which the oil is at the bottom and the saturated network is the clear phase.

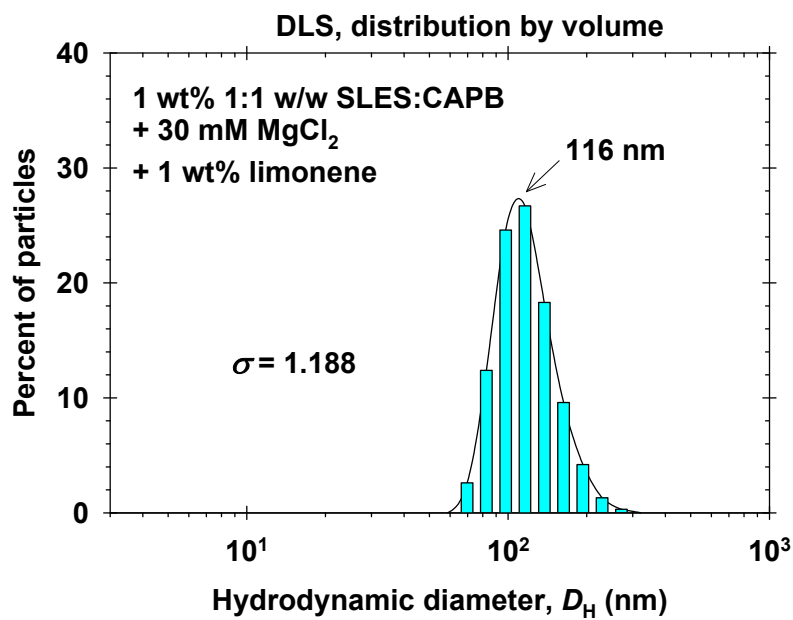

**Fig. S7.** Dynamic light scattering (DLS): Size distribution of the drops in a nanoemulsion obtained by the addition of 1 wt% limonene to the multiconnected micellar phase separated from a solution of total concentration 1 wt% 1:1 w/w CAPB:SLES-1EO + 30 mM $MgCl_2$. The size distribution is very close to the lognormal one; the most probable (median) drop diameter is $\bar{D}_H = 116$ nm; $\sigma = 1.188$ is the standard deviation on the log scale; 68.3% of the drop diameters belong to the interval between $\bar{D}_H / \sigma$ and $\bar{D}_H \sigma$, that is between 98 and 138 nm. The size distribution obtained by DLS is in very good agreement with the one obtained by laser diffraction – see Fig. 8c in the main text.

Comment on the SAXS spectrum in Fig. S4

By structure, the multiconnected micellar network resembles the disordered *sponge phase* (L3) observed in microemulsions and phospholipid self-assemblies [2,6]. Just like the sponge phase, the micellar network displays a single pronounced peak, which is centered at $q = 0.483$ Å$^{-1}$ (Fig. S4) and appears on a decaying background curve. Such a peak is typical for fluid isotropic phases [2]. In view of Eq. (1) in Ref. [2], this peak corresponds to a reference distance $L = 2\pi/q = 1.30$ nm, which is very close to the expected radius of the cylindrical micelle; see e.g. Fig. 8a in Ref. [7]. In Ref. [2], the peak corresponds to the thickness of the phospholipid bilayer, ca. 3.2 nm. The obtained values are related to the fact that the cylindrical micelles (here) and phospholipid bilayers (Ref. [2]) are, respectively, linear and planar structures; see Ref. [8].